\documentclass[11pt,twoside]{article}
\usepackage{asp2006}
\usepackage{adassconf}
\usepackage{graphicx}
\begin{document}

\paperID{P89}

\title{The Virtual Observatory Powered PhD Thesis}
\markboth{Zolotukhin}{The VO-powered PhD Thesis}

\author{Ivan Yu. Zolotukhin}
\affil{Sternberg Astronomical Institute, Moscow State University, Universitetskij pr., 13, 119992, Moscow, Russia}

\contact{Ivan Zolotukhin}
\email{iz@sai.msu.ru}

\paindex{Zolotukhin, I.~Yu.}

\keywords{astronomy!Galactic, astronomy!extragalactic, astronomy!IR, catalogs!cross-match, catalogs!individual}

\begin{abstract}
The Virtual Observatory has reached sufficient maturity for its routine scientific exploitation by astronomers. To prove this statement, here I present a brief description of the complete VO-powered PhD thesis entitled ``Galactic and extragalactic research with modern surveys and the Virtual Observatory'' comprising 4 science cases covering various aspects of astrophysical research. These comprize: (1) homogeneous search and measurement of main physical parameters of Galactic open star clusters in huge multi-band photometric surveys; (2) study of optical-to-NIR galaxy colors using a large homogeneous dataset including spectroscopy and photometry from SDSS and UKIDSS; (3) study of faint low-mass X-ray binary population in modern observational archives; (4) search for optical counterparts of unidentified X-ray objects with large positional uncertainties in the Galactic Plane. All these projects make heavy use of the VO technologies and tools and would not be achievable without them. So refereed papers published in the frame of this thesis can undoubtedly be added to the growing list of VO-based research works.
\end{abstract}

\section{Introduction}

Following several years of intensive technological development, Virtual Observatory is now starting to be used frequently by wider research community for diverse range of scientific studies. We observe emerging projects that go further beyond data mining (see e.~g.\ recent papers by Chilingarian et al.\ (2009); D'Abrusco, Longo, \& Walton (2009); Valdivielso et al.\ (2009) and Chilingarian (2009) for the review) making essential use of advanced VO technologies. In this context it was possible to prepare the first (up to our knowledge) PhD thesis explicitly based on the VO methods and technologies. The work entitled ``Galactic and extragalactic research with modern surveys and the Virtual Observatory'' was successfully defended in October 2009 in the Moscow State University receiving essentially positive feedback among Russian professional astronomical community. Different astrophysical problems were studied using the single research approach behind them: data archives and VO resources are to be analyzed thoroughly and, combined with dedicated observations on large telescopes (if necessary), they can provide deep insights into wide range of astrophysical problems. The formal process adopted in the Moscow University does not allow writing thesis in English, so I outline its main results with some VO advocacy in this paper.

\section{Open Clusters Study}

\begin{figure}
\plottwo{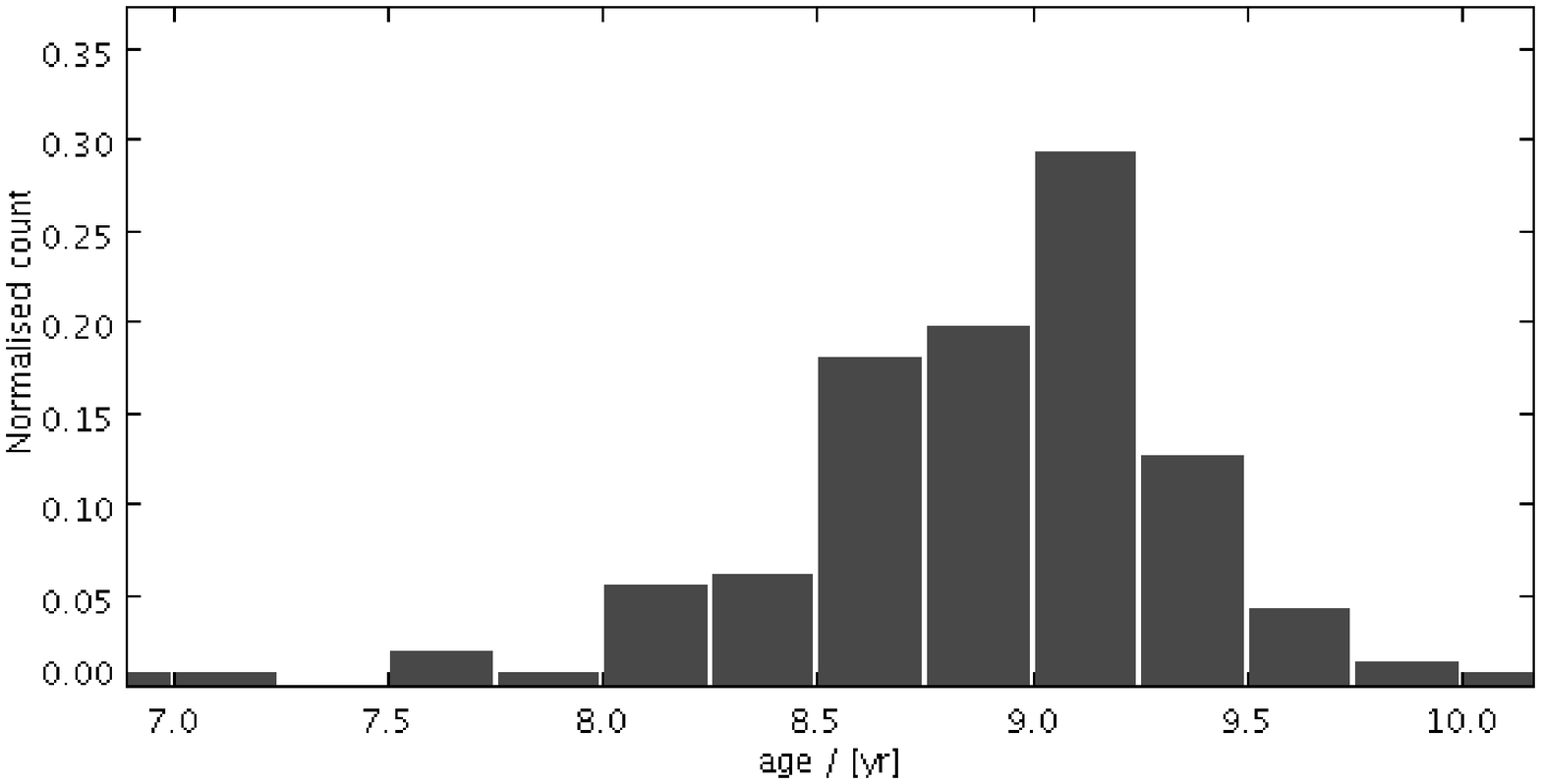}{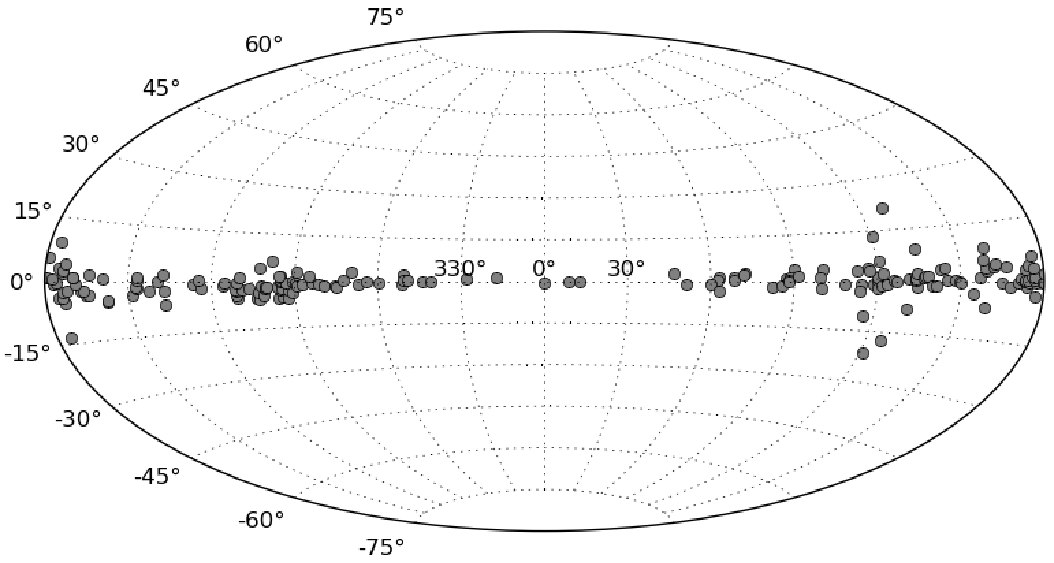}
\caption{Left panel: age distribution of the new clusters discovered in the frame of this study; right panel: distribution of the new clusters on a sky in Galactic coordinates.} \label{P89-fig-1}
\end{figure}

We developed an automated method to search for open clusters in large multi-band surveys (Koposov, Glushkova \& Zolotukhin 2008). The algorithm finds star density peaks and tests their color-magnitude diagrams fitting an isochrone there. If the procedure converges, we consider peak to be a real cluster and at the same time get an estimate of the age, distance and color excess for it. Using standardized VO access methods to the large catalog collection at Sternberg Astronomical Institute (Koposov et al.\ 2007) we applied our algorithm to the 2MASS data in the stripe $|b| < 24$ degrees along the Galactic Plane and found 168 new open clusters (see their distribution by age and on sky in Fig.~\ref{P89-fig-1}) increasing by ~10\% the information about this important subsystem of the Galaxy (Glushkova et al.\ 2009). The results of the ongoing study, SAI Open Clusters Catalog, are presented in a VO-ready form at the project \htmladdnormallinkfoot{web-site}{http://ocl.sai.msu.ru}. Apart from convenient presentation and necessary VO access interfaces, the site provides advanced VO experience by a Java applet providing possibilities of the direct interaction (launch and manipulation) of client VO applications right from the web browser (Zolotukhin \& Chilingarian 2008). There is an ongoing effort to observe these clusters with 1-m class telescopes to confirm reliability of our method. This study showcases the potential of all-sky surveys and the Virtual Observatory for homogeneous studies which is far from being exhausted.

\section{Optical and IR Identification of X-ray Binaries}

\begin{figure}
\plottwo{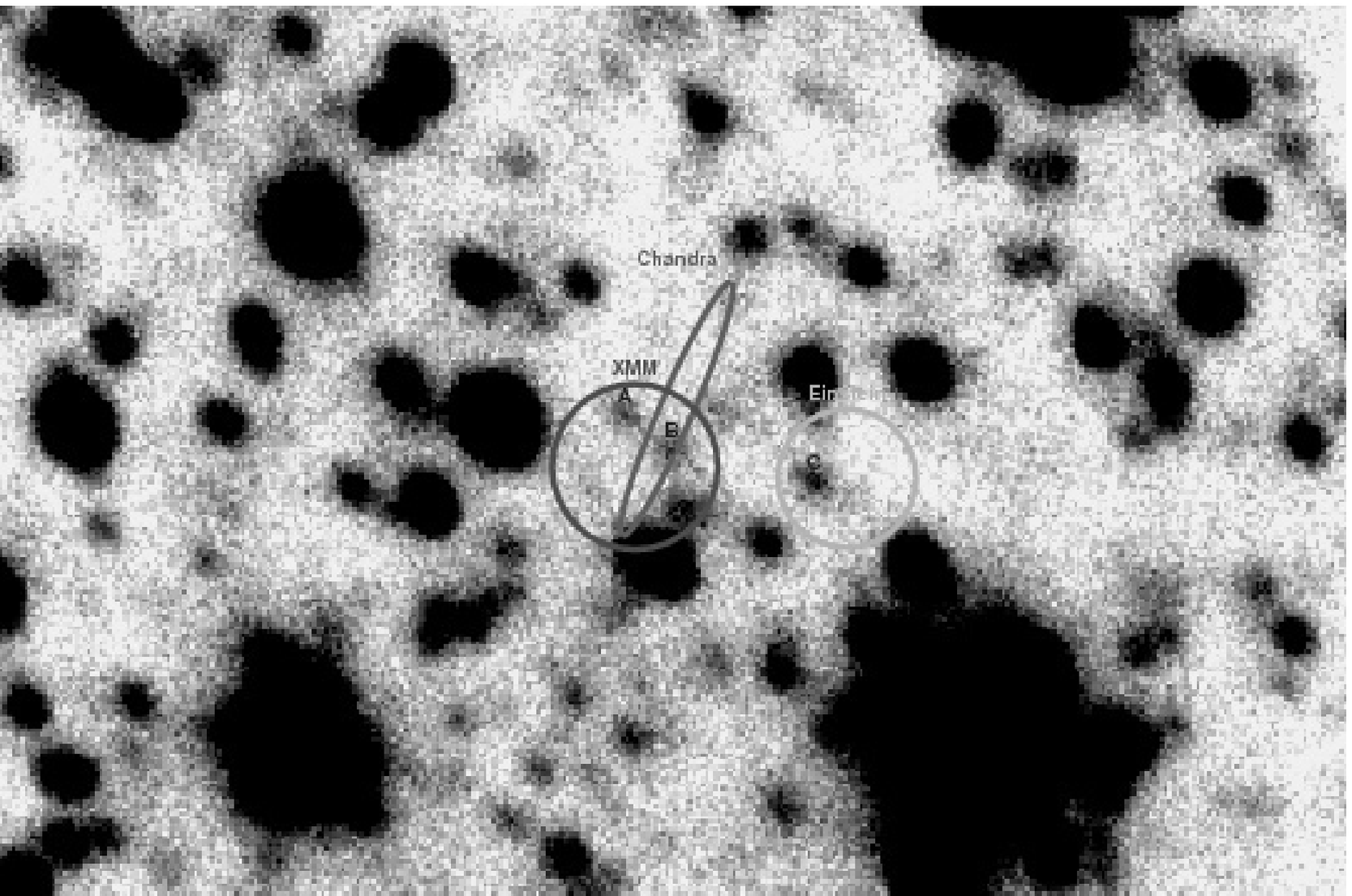}{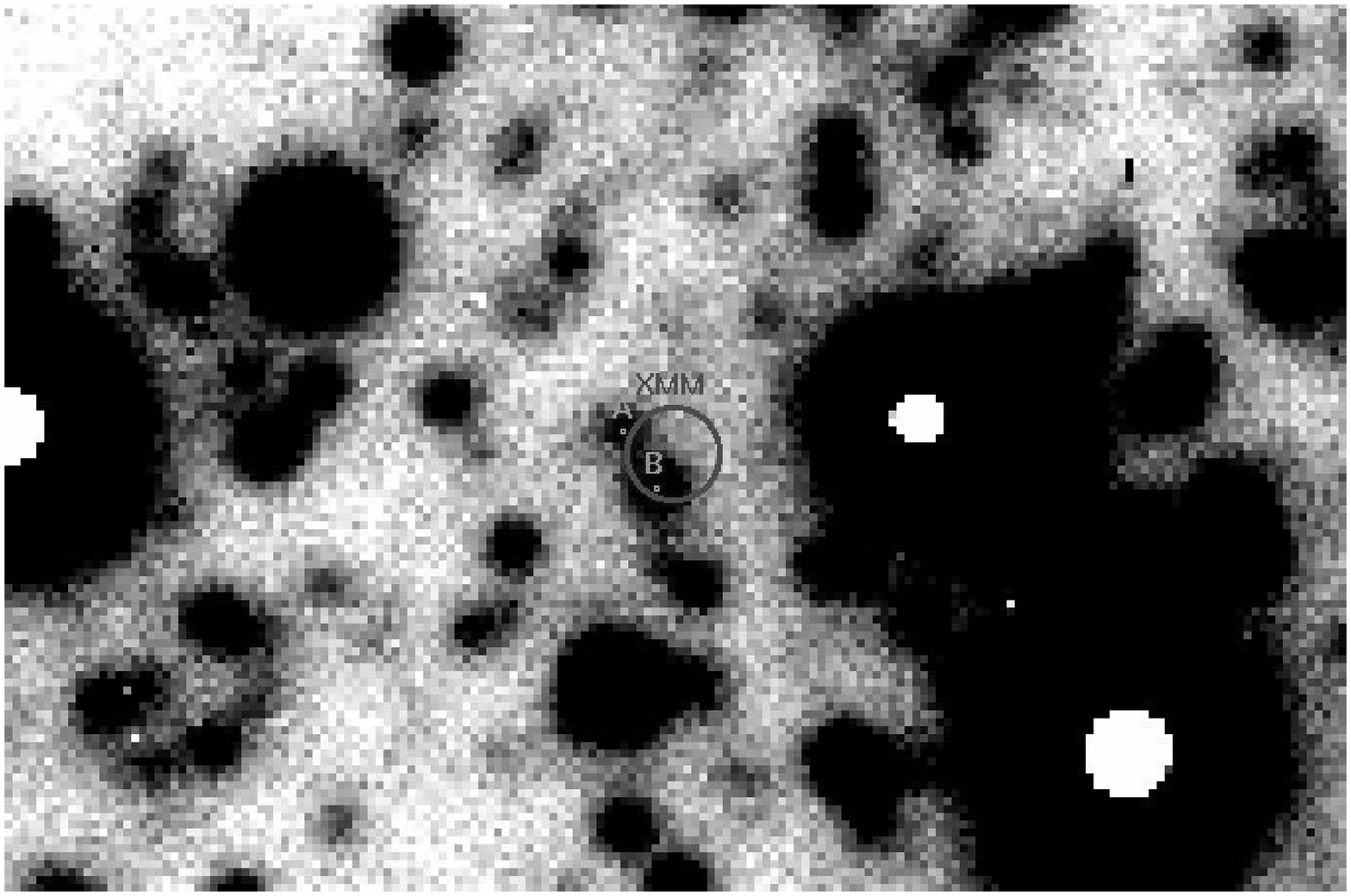}
\caption{Fields of X-ray binaries with {\it Chandra}/{\it XMM} positional uncertainties overplotted. Data were taken from ESO archive. Left panel: field of 4U1323-619 observed in filter $K_s$ by ESO 3.6-m NTT with SOFI detector (Zolotukhin, Revnivtsev, \& Shakura 2010); right panel: field of IGR J17254-3257 observed in filter $I$ by ESO 3.6-m NTT with EMMI detector (Zolotukhin 2009).} \label{P89-fig-2}
\end{figure}

In order to constrain sizes of some X-ray binaries we undertake photometrical measurements of those which do not possess reliable optical or infrared counterparts. Given the X-ray luminosity and the observed optical/NIR flux it is possible to estimate accretion disk size and overall system extent. This allows us to discover new ultracompact low mass X-ray binaries which is a small and poorly-studied population of the Galactic objects expected to emit gravitational waves. We used VO data discovery methods to check large existing observational collections for the data on potential candidates and found a number of observed sources of interest without published measurements. Three low mass X-ray binaries, 4U1323-619, IGR J17254-3257 and SLX 1735-269 (see two of them in Fig.~\ref{P89-fig-2}) were identified in archival data then (ESO, {\it Chandra} and {\it XMM} archives used). In case of 4U1323-619 there was a mis-identification based on {\it Einstein} X-ray observations with an underestimated error radius. Even without reliable identification, one can establish upper limit on brightness of a source that immediately implies upper limit on its disk size. This study demonstrates the importance of VO methods of data discovery since existing rich observational archives may contain comprehensive datasets which can give us clues about nature of poorly-known binary populations.

\section{Optical-to-NIR Colors of Nearby Galaxies}

We cross-identified a spectral sample of SDSS DR7 galaxies lying at $0.03 < z < 0.6$ with UKIDSS DR5 data, and fitted their spectra with simple stellar population models, thus constructing a catalog of 200k galaxies with $ugriz$ and $YJHK$ photometry measurements, redshifts, spectra, ages and metallicities. This allowed us to determine simple yet precise analytical approximations of $k$-corrections having great practical value for extragalactic research (Chilingarian, Melchior, \& Zolotukhin 2010). The traditional $k$-correction computation techniques based on the SED-fitting, require the multi-color photometry, which often is not available. Our approach allows one to compute rest-frame galaxy magnitudes of the same quality using a minimal set of observables including only two photometric points and a redshift.

\section{X-ray sources in the Galactic Plane}

\begin{figure}
\epsscale{0.30}
\plotone{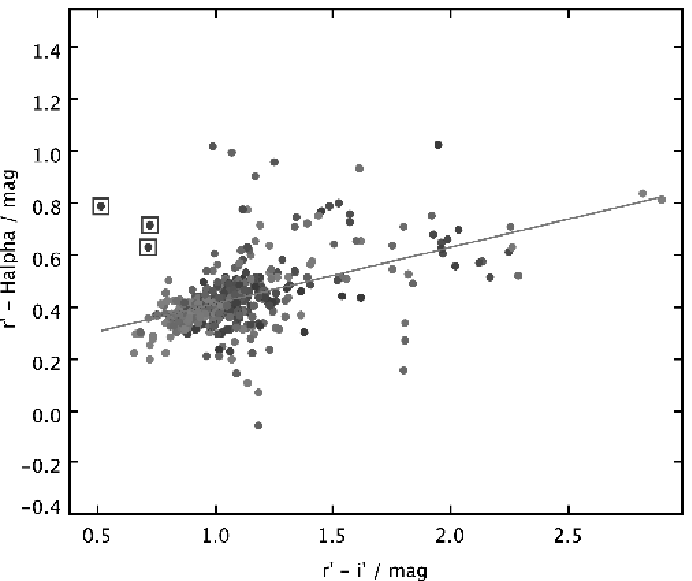}
\epsscale{0.40}
\plotone{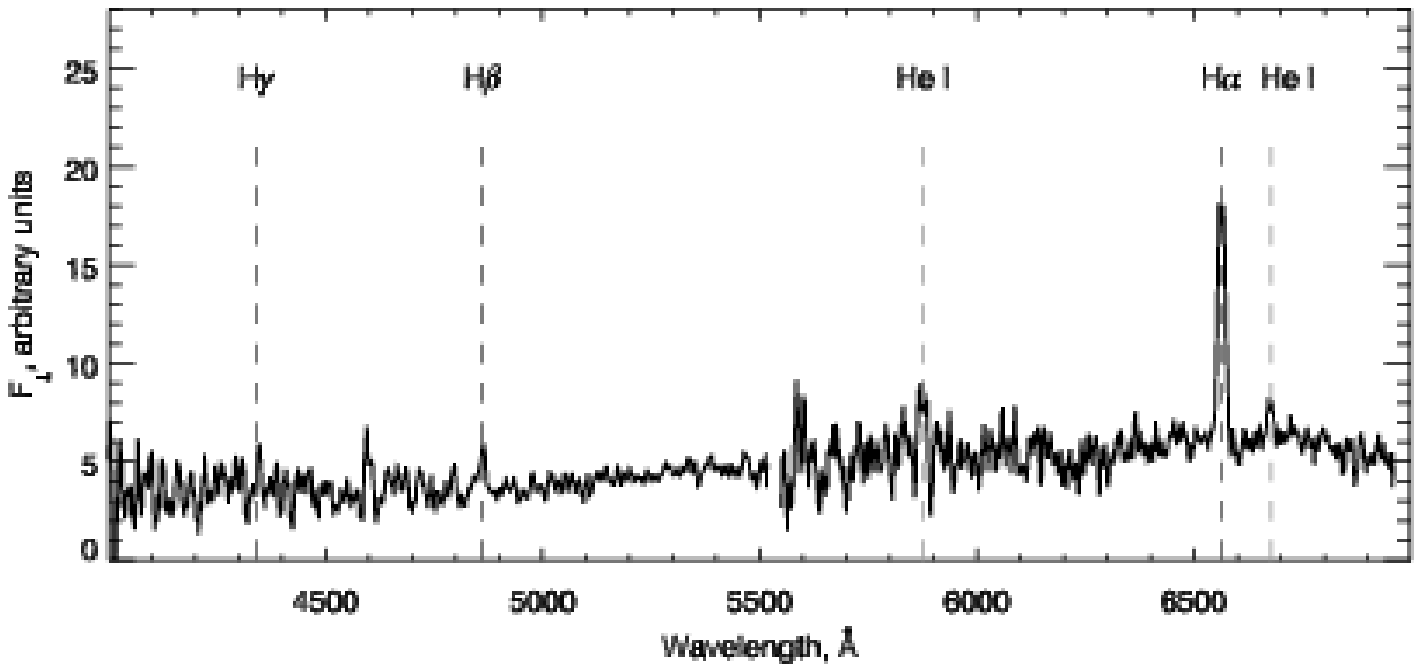}
\caption{AX J194939+2631 identification. Left panel: color-color diagram of IPHAS detections in 1\arcmin ~field around ASCA coordinates of the source; three measurements of the potential counterpart demonstrating H$\alpha$ emission excess are shown in squares. Right panel: optical spectrum of the source obtained at 3.5-m Calar Alto telescope, time granted in DDT quota on a basis of the input from the VO. The spectrum confirms the source to be X-ray active cataclysmic variable.} \label{P89-fig-3}
\end{figure}

Using IPHAS (INT Photometric H$\alpha$ Survey) we identified the X-ray source AX J194939+2631 from ASCA Galactic Plane Survey that initially had 1\arcmin ~positional uncertainty which contained hundreds of objects in modern surveys. By imposing highly selective criteria it becomes possible to distinguish between background stars and the object of interest if it is included in the survey. Assuming that some set of unidentified ASCA X-ray sources should have hydrogen-rich accretion disks and therefore exhibit H$\alpha$ emission excess, we used the method similar to color-color diagram analysis proposed by Witham et al.\ (2006) (see left panel in Fig.~\ref{P89-fig-3}) to extract emitters in fields of interest. In bright case of AX J194939+2631 we confirmed the nature of the most prominent H$\alpha$ emitter in the field by means of optical spectroscopy (see right panel in Fig.~\ref{P89-fig-3}). This use-case is listed as an example of the comprehensive VO examination in order to identify X-ray source with challenging positional uncertainty.

\acknowledgments
Author wish to thank ADASS organizing committee for the financial support
provided. Travel was also supported via RFBR grant 09-02-09603.

\end{document}